\documentclass[pre,aps,reprint,floats,showkeys,superscriptaddress]{revtex4-2}
\usepackage{amssymb,amsmath}
\usepackage{graphicx}

\makeatletter
\renewcommand\frontmatter@abstractwidth{\dimexpr\textwidth-1in\relax}
\makeatother

\usepackage{etoolbox}
\patchcmd{\section}
  {\centering}
  {\raggedright}
  {}
  {}
\patchcmd{\subsection}
  {\centering}
  {\raggedright}
  {}
  {}

\begin{document}

\title{Thermodynamics of interacting hard rods on a lattice} 

\author{Tounsi Benmessabih}
\email{tounsi.benmessabih@univ-mascara.dz}
\affiliation{Department of Physics, University Mustapha Stambouli of Mascara, 29000 Mascara, Algeria}
\author{Benaoumeur Bakhti}
\email[Corresponding author: ]{bbakhti@uni-osnabrueck.de}
\affiliation{Department of Physics, University Mustapha Stambouli of Mascara, 29000 Mascara, Algeria}
\author{M. Reda Chellali}
\email[Corresponding author: ]{mohammed.chellali@kit.edu}
\affiliation{Institute of Nanotechnology, Karlsruhe Institute of Technology, Hermann-von-Helmholtz-Platz 1, 76344 Eggenstein-Leopoldshafen, Germany}


\begin{abstract}
\textbf{Abstract}\\
We present an exact derivation of the isobaric partition function of lattice hard
rods with arbitrary nearest neighbor and long rang interactions. Free energy and all thermodynamics functions are derived accordingly and they are written in a form that is suitable for numerical implementation. As an application, we have considered lattice rods with pure hard-core interactions, rods with long-range gravitational attraction, and finally a charged hard rods with charged boundaries (Bose gas), a model that is relevant for studying several phenomena such as charge regulation, ionic liquids near charged interfaces, and an array of charged smectic layers or lipid multilayers. In all cases, thermodynamic analyses have been done numerically using the Broyden algorithm.
\end{abstract}
\keywords{Hard rods; long range interactions; lattice systems}

\maketitle

\section{Introduction}\label{sec:sec_1}
The physics of one-dimensional structures attracted much interest 
both in theory and in experiment due to its relevant for nanotechnology \cite{Colinge/Greer:2016,Xu/etal:2011,Dreyfus/etal:2009}.
With the emergence of the manipulation and high resolution
imaging instruments such as the scanning tunneling
microscopy, scanning tunneling spectroscopy
 and atom probe tomography \cite{Chellali/etal:2020,Chellali/etal:2019,Chellali/etal:2016,Chellali/etal:2012}, it is possible in current experiments to
manipulate one-dimensional chains of atoms or molecules
 \cite{Flohr/etal:2012,Iancu/etal:2009,Zhou/etal:2008}. These chains
have been shown to have a unique and fascinating
properties which do not exist in two or three dimensions \cite{Dubin:1997,Prestipino/Giaquinta:2003, Prestipino:2003}.

One dimensional hard rods played a central role in the
development of classical density functional theory 
\cite{Bakhti/etal:2012,Bakhti/etal:2013,Bakhti:2013a,Bakhti/etal:2014,Bakhti/etal:2015c,Girardeau/Astrakharchik:2012}
and it is a paradigm model for the statistical mechanics
of an extended molecules or polymers \cite{Harnau/Dietrich:2002,Harnau/Dietrich:2007}. It has been extensively
studied and many exact results have been derived. Beside their mathematical 
interest, the exactly solvable models in one dimension are important to test the quality of the simulation experiments and they provide
an insight to the structure of the theory in three-dimension such
as the fundamental measure theory \cite{Rosenfeld:1989, Lafuente/Cuesta:2002b,Tarazona/etal:2008} which has been inspired in its 
construction by the Percus free energy for hard rods in an arbitrary 
external field \cite{Percus:1976,Robledo/Varea:1981}. In addition, if the external potential is uni-directional, the $2D$ and the $3D$ systems are reduced to effective $1D$ systems in which the $1D$ effective density is the planar average of density of the higher dimensional systems. 

Despite its important role in the development of the classical theory of fluids, only a few results are available for hard rods with interaction beyond the hard-core exclusion \cite{Percus:1989,Vanderlick/etal:1989, Brannock/Percus:1996,Percus:1997,Mortazavifar/Oettel:2017}, and the thermodynamics of these systems have been very rarely invoked in the literature \cite{Harnau/Dietrich:2007}. Even more, results for lattice systems are scarce. The lattice description of classical fluids \cite{Lafuente/Cuesta,Bakhti/etal:2018} are shown to be simple and very useful in many cases such DNA denaturation \cite{Azbel:1979}, fluids in porous media \cite{Kierlik/etal:2001}, glasses \cite{Biroli/Mezard:2001} and roughening phenomena \cite{Chui/Weeks:1981}. Thus, it is our aim in this paper to set up a general formalism for treating lattice hard rods with interactions in the isobaric ensemble \cite{Bishop/Boonstra:1983}  that it suitable for numerical implementation.  Then, detailed numerical studies of three cases will be presented. The paper is organized as follow: we consider first the general problem of hard rods with an arbitrary nearest neighbor and long-range interactions. Then we make applications the pure hard rods system, hard rods with gravitational attraction, and Bose gas of hard rods which plays a dominant role in the formation of many one-dimensional structures. We conclude with some outlooks for future work. 

\section{General Formalism}\label{sec:sec_2}
Consider a lattice system of $N$ hard rods of equal lengths $\sigma$ in a unit of lattice parameter $a$ supposed to be one here. The rods are distributed on a one-dimensional lattice of $L$ sites. The continuum limit of this system can be recovered by letting $\sigma\rightarrow\infty$ and $a\rightarrow 0$ and keeping $\sigma a$ fixed.
The position of the rod $i$ ($x_i$) is labeled by the position of its left end so that if a rod is located at a position $i$, no other rod can be placed in the position $i+1,\ldots,i+\sigma$. Without loss of generality, we suppose that the rods are ordered such that $0 < z_1 < z_2,\ldots, < z_N <L$. This fixed spatial sequence of the rods is a consequence of both unidimensionality and the hardness of particles. Free boundary conditions are considered here. The formalism works also for fixed boundary conditions in which two particles are fixed at the first and last sites. Besides the hard-core exclusion, the rods interact through an arbitrary two-body interaction and they are subject to uniform external field and temperature. To calculate the exact partition function of this system we adapted a procedure suggested in Ref \cite{Bishop/Boonstra:1983} to calculate the partition function of the continuum system of hard rods. Let us first consider a system of hard rods with nearest-neighbor interaction. The canonical partition function can be written as \cite{comm:kinetic_part_function}
\begin{equation}\label{eq:47}
\mathcal{Z}=\sum\limits_{x_N=0}^{L}\sum\limits_{x_{N-1}=0}^{x_N}\ldots \sum\limits_{x_1=0}^{x_2}t(x_1)t(x_2-x_1)\ldots t(L-x_N)
\end{equation}
where $x_i$ is a discrete position of rod $i$ on the lattice ($x_i$ is expressed in a unit of the lattice parameter $a$ which we have omitted here for simplicity). The factor $N!$ in the canonical partition function is canceled by the $N!$ permutation of the rods ordering. The function $t(z)$ is given by
\begin{equation}\label{eq:48}
t(z)=e^{-\beta \phi(z)}
\end{equation}
$\phi(z)$ is an arbitrary nearest neighbor interaction (that includes also the hard-core interaction) and $z$ is the discrete separating distance between the rods ($z_i=x_{i+1}-x_i$). We can prove using the convolution theorem of the Laplace transform that the Laplace transform of the partition function Eq.~(\ref{eq:47}) is given by
\begin{equation}\label{eq:54}
\mathcal{\tilde{Z}}_N=(\tilde{t})^{N+1}
\end{equation}
where
\begin{equation}\label{eq:52}
\tilde{t}(s)=\sum\limits_{z=0}^{\infty}e^{-\beta pz}t(z)
\end{equation}
where $s$ is the conjugate variable to $z$. For application in statistical mechanics, $s$ represents the pressure $p$ of the system, and $\mathcal{\tilde{Z}}$ is the partition function in the isobaric ensemble. The canonical partition function is the inverse Laplace transform of $\mathcal{\tilde{Z}}$.
Despite it is not always possible to calculate the inverse Laplace transform of $\mathcal{\tilde{Z}}$,
the latter itself is important because its thermodynamic average agrees to $1/N$ with those of the canonical partition function \cite{Lebowitz/Percus:1961}. The problem of calculating the partition function is reduced to calculate the function $\tilde{t}$
and all thermodynamic functions can be inferred from the latter via differentiation.
\section{long range interaction}\label{sec:sec_3}
The precedent calculations can be generalized to the case where the interaction is not limited to the nearest-neighbor interaction. In this case, the potential energy of each particle results from the interaction with all other particles of the system. We present here a general formalism to deal with the long-range interaction and we present two specific examples in the next section. The total potential energy is given by
\begin{align}
H = \sum\limits_{i<j}\phi_{i,j}
\end{align}
where $\phi_{i,j}=\phi(x_i,x_j)$ is the interparticle interaction, $x_i$ the position on the lattice of rod $i$. The total interaction potential between particle $i$ and the rest of the system can be written as
\begin{align}\label{eq:nn_int}
\phi_{i}(x_1,x_2,\ldots,x_N) = \sum\limits_{j\neq i}^{N} \phi_{i,j}
\end{align}
If the pair interactions depend only on the separating distance between the rods, .\i.e $\phi(x_i,x_j)=\phi(|x_j-x_i|)$, then the coordinates $\{x_1,x_2,\ldots,x_N\}$ can be mapped to a new coordinates representing the separating distance between the rods in addition to the center of mass, \i.e. $\{x_G,z_1,z_2,\ldots,z_{N-1}\}$ where $z_i=x_{i+1}-x_i$ and $x_G=\sum_{i=1}^Nx_i/N$. In this case, we can write $\phi(z)=\phi(|x_j-x_i|)$. As will be shown in the examples bellow, if the interactions depend only on the separating distance between the particles, the total potential energy can be written as
\begin{align}
\phi(x_1,x_2,\ldots,x_N)&=\sum_{i=1}^N\phi_{i}(x_1,x_2,\ldots,x_N)\nonumber\\
&=\sum_{i=1}^{N}\phi_i(z_1,z_2,\ldots,z_{N-1})
\end{align}
where $z_i=x_{i+1}-x_i$. For hard rods of length $\sigma$, the 
interaction energy Eq.~(\ref{eq:nn_int}) manifests in the hard core exclusion which prevents overlapping of particles. Because the later extends over the rod length $\sigma$,
the function $\tilde{t}$ is given by
\begin{align}\label{eq:4.84}
\tilde{t}_i(p,T)=\sum\limits_{z_i=\sigma}^{\infty}\omega_i(z_i).
\end{align}
Let us introduce the following notations
\begin{equation}
\omega_i(z) = e^{-\beta (pz+\phi_i(z))}.
\end{equation}
The free energy is given by
\begin{align}\label{eq:g}
\beta G=-\sum_{i=0}^{N}\ln\tilde{t}_i,
\end{align}
from which all thermodynamic functions can be inferred. Density $\rho$, entropy $S$, internal energy $U$, heat capacity $C_p$, isothermal compressibility $\chi_T$ and thermal expansion coefficient $\alpha_V$ are given by
\begin{align}\label{eq:es}
\frac{1}{\rho}\!=\! \frac{1}{N}\left(\frac{\partial G}{\partial p}\right)_{\!\! N,T}\!=\! \frac{1}{N}\sum_{i=0}^{N}\frac{1}{\tilde{t}_i}\frac{\partial\tilde{t}_i}{\partial p},
\end{align}
\begin{align}\label{eq:s}
\frac{S}{k_B}\!=\!\left(\frac{\partial G}{\partial T}\right)_{\!\! p,N}\!=\!\sum_{i=0}^{N}\left[\ln\tilde{t}_i - \frac{\beta}{\tilde{t}_i}\frac{\partial\tilde{t}_i}{\partial\beta}\right],
\end{align}

\begin{align}\label{eq:u}
U \!=\! G+TS-pL \!=\! -\sum_{i=0}^{N}\!\frac{1}{\tilde{w}_i}\left[\frac{\partial\tilde{t}_i}{\partial\beta}+p\frac{\partial\tilde{t}_i}{\partial p}\right],
\end{align}

\begin{equation}\label{eq:cp}
\frac{C_p}{k_B}\!=\! T\left(\frac{\partial S}{\partial T}\right)_{\!\! p}\!=\! \beta^2\!\sum_{i=0}^{N}\!\left[\!\frac{1}{\tilde{t}_i}\frac{\partial^{2}\tilde{t}_i}{\partial\beta^{2}} - \left(\!\frac{1}{\tilde{t}_i}\frac{\partial\tilde{t}_i}{\partial\beta}\!\right)^{2}\right],
\end{equation}

\begin{equation}\label{chi}
\chi_T \!=\! -\frac{1}{L}\!\left(\!\frac{\partial L}{\partial p}\!\right)_{\!\! T}\!\!=\! \frac{\rho}{N}\!\sum_{i=0}^{N}\!\left[\!\left(\!\frac{1}{\tilde{t}_i}\frac{\partial\tilde{t}_i}{\partial p}\!\right)^{2}\!\! -\frac{1}{\tilde{t}_i}\frac{\partial^{2}\tilde{t}_i}{\partial p^{2}}\!\right],
\end{equation}

\begin{equation}\label{eq:alpha}
\frac{\alpha_V}{k_B}\!=\! -\frac{1}{L}\!\left(\!\frac{\partial L}{\partial T}\!\right)_{\!\! p} \!\!=\! \frac{\beta^2\rho}{N}\!\sum_{i=0}^{N}\!\frac{1}{\tilde{t}_i}\!\!\left[\frac{\partial^{2}\tilde{t}_i}{\partial\beta\partial p}\!-\!\frac{1}{\tilde{t}_i}\frac{\partial\tilde{t}_i}{\partial\beta}\frac{\partial\tilde{t}_i}{\partial p}\!\right].
\end{equation}
where $L=N/\rho$ is the volume. The problem of calculating the thermodynamics functions is reduced to calculate the function $\tilde{t}_i$ and its derivative with respect to pressure $p$ and inverse temperature $\beta$. The previous general formalism for treating the nearest neighbor and long-range interaction can be treated numerically for systems with any kind of interaction. the advantage of the formalism presented in this work is that it is easier for numerical treatment compared to other approaches. In some cases such as hard rods with only hard-core exclusion, all thermodynamic functions Eqs.~(\ref{eq:g})-(\ref{eq:alpha}) can be evaluated analytically and explicitly as a function of the control parameters $N$, $p$ and $T$. For long-range interactions, variation of thermodynamic function versus density can be still evaluated with a direct numerical calculation. To get the variation with respect to temperature, the numerical problem reduces to solve a system of nonlinear equations   Eqs.~(\ref{eq:g})-(\ref{eq:alpha}) and we have done this using the Broyden algorithm.
\section{Applications}
In this section, we consider three cases that are relevant for many applications such as sedimentation of effective one-dimensional colloids under gravity \cite{Bakhti/Muller:2021} or ordering transition in a one-dimensional system with Coulomb interactions \cite{King/etal:2001}.

\subsection{Hard rods}
\begin{figure}[htb]
  \begin{center}
 \includegraphics[width=42mm]{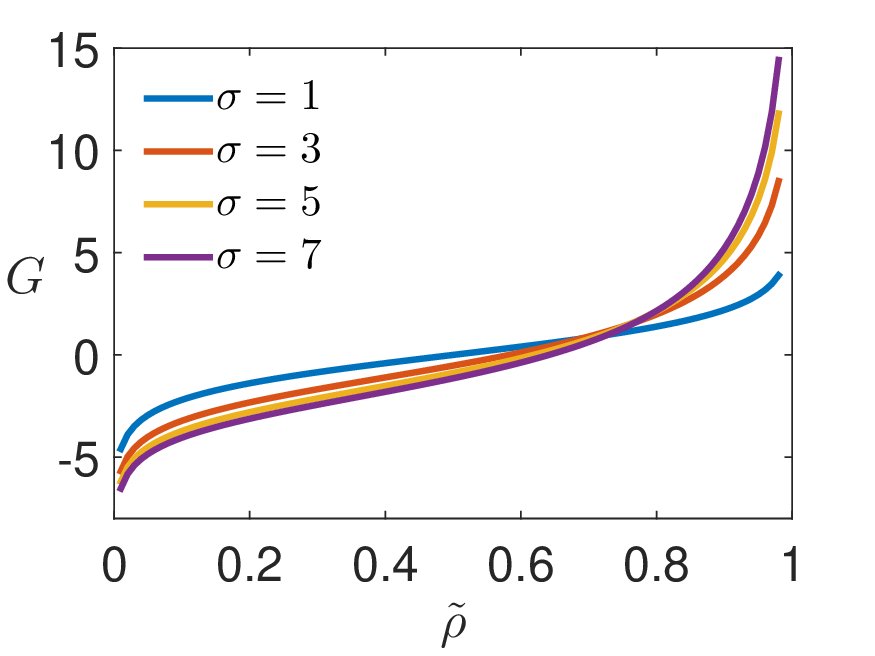}
 \includegraphics[width=42mm]{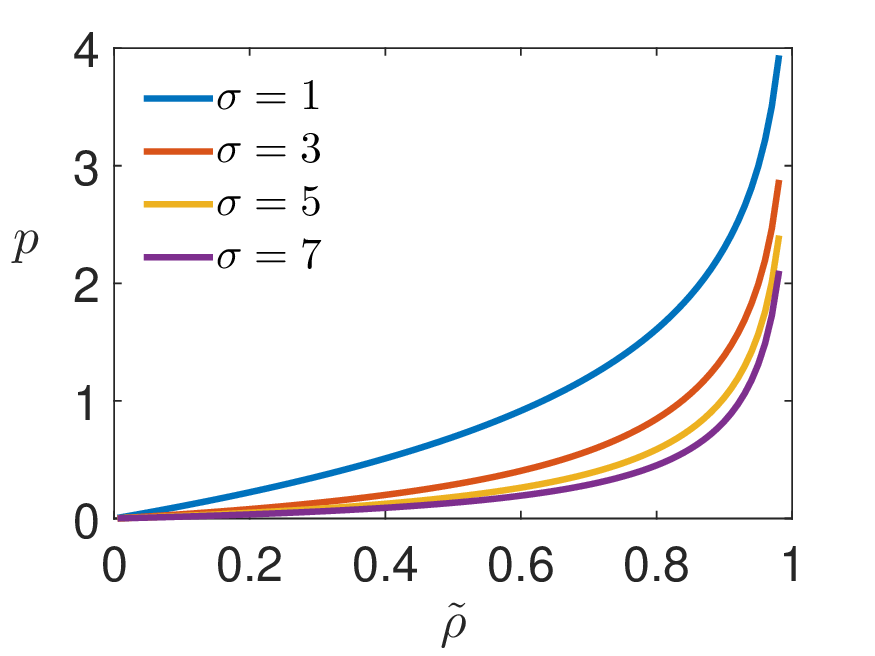}\\
 \includegraphics[width=42mm]{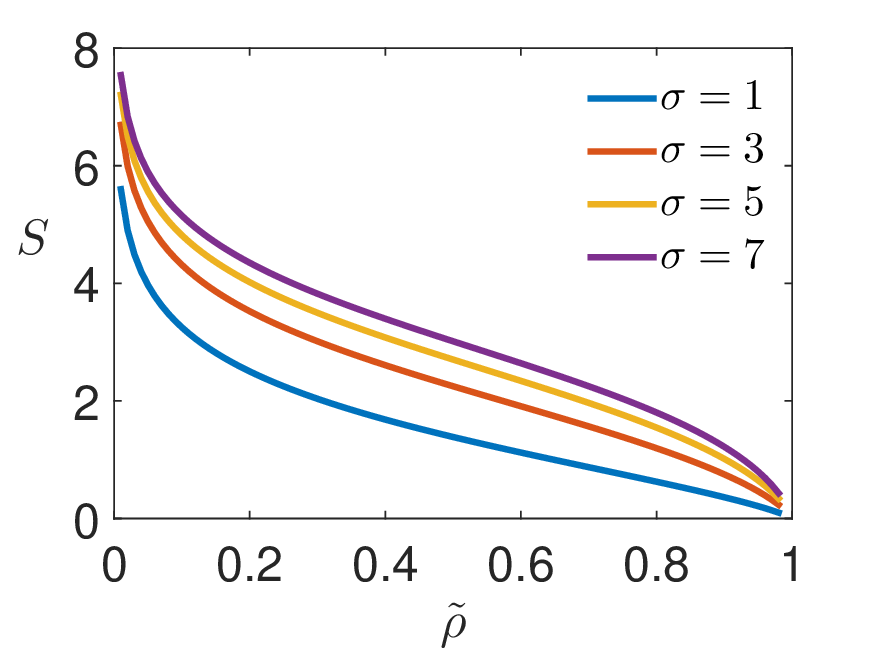}
 \includegraphics[width=42mm]{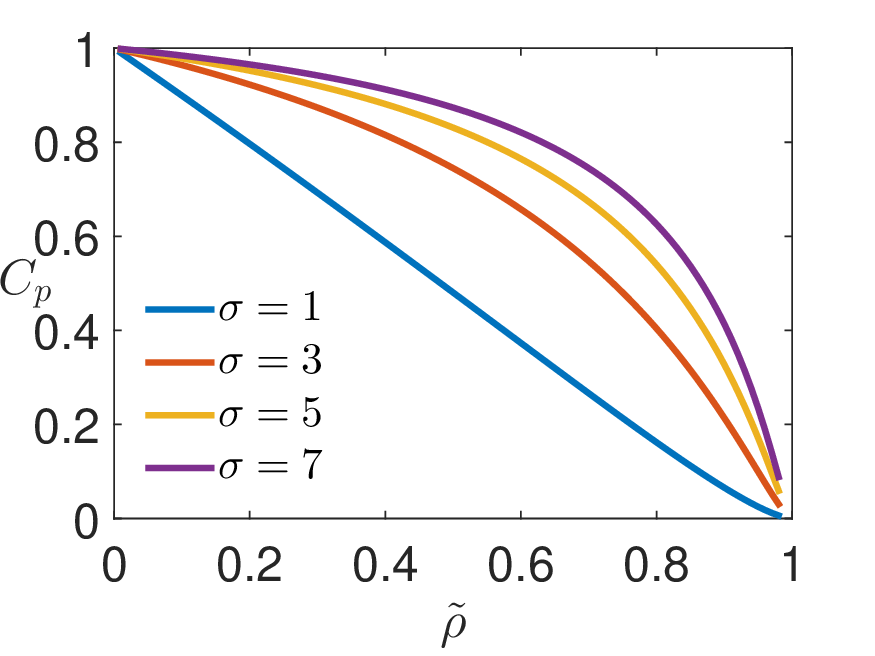}\\
 \includegraphics[width=42mm]{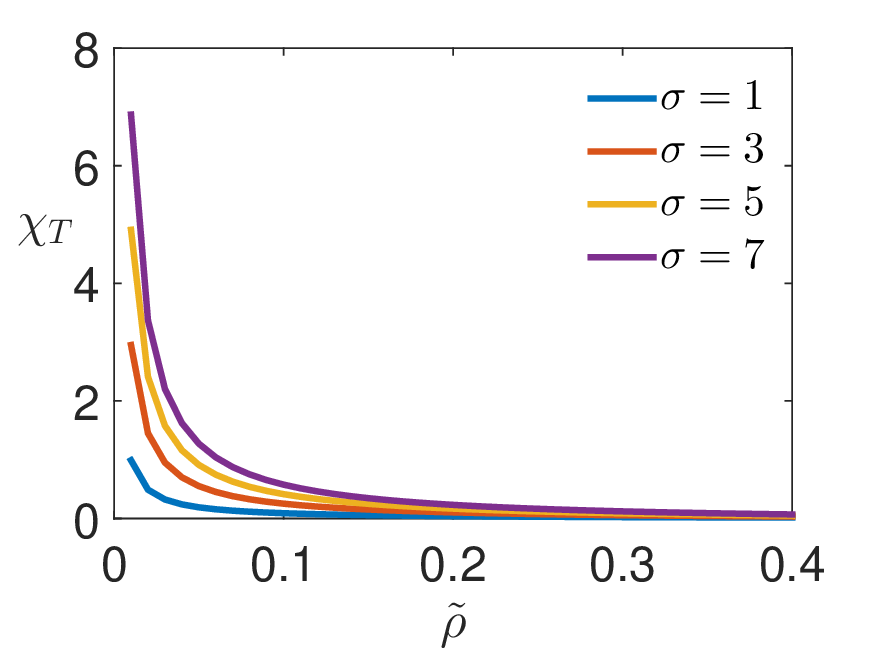}
 \includegraphics[width=42mm]{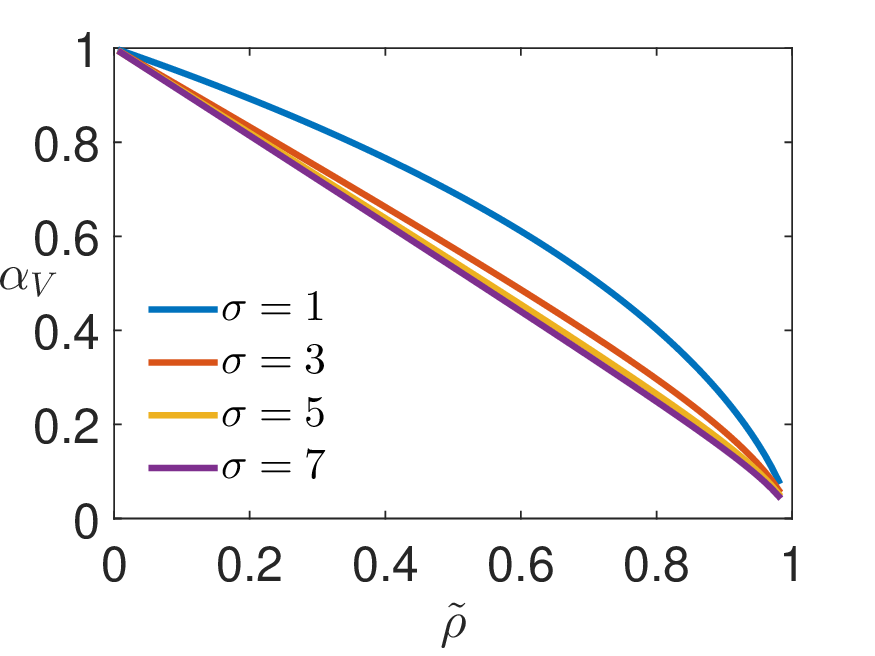}
\end{center}
\caption{Free energy $G$, equations of state $p(\tilde{\rho})$, entropy $S$, heat capacity $C_p$, isothermal compressibility $\chi_T$ and coefficient of thermal expansion $\alpha_V$ of hard rods on with different lengths. All quantities are scaled by the number of particles $N$}
  \label{fig:fig1}
\end{figure}

System of pure hard rods has been investigated in the literature both in continuum \cite{Jang/etal:2017,Champion:2015,Kierlik/Rosinberg:1992,Zhang:1991,Honnell/Hal:1990,Percus:1976} and on  a lattice \citep{Bakhti/etal:2012,Bakhti/etal:2013,Bakhti:2013a,Bakhti/etal:2014,Bakhti/etal:2015c} for mono and multicomponent cases. However, to the best of our knowledge, there is no detailed thermodynamic analysis of this system. Here we show how the general formalism can be adapted to this system and present numerical analysis of all the thermodynamics functions, which could be useful for testing computer simulations. Let us consider the case of hard rods with length $\sigma$ (in a unit of lattice constant $a$). The inter-atomic coupling is given by
\begin{align}\label{eq:64}
\phi(z)=\left\lbrace\begin{array}{ll}
\infty  & \mbox{ for $z < \sigma$},\\
0       & \mbox{for $z \ge \sigma$}.
\end{array}\right.
\end{align}
The kinetic energy produces a constant term that is not relevant for thermodynamics \cite{comm:kinetic_part_function}. Inserting the interaction energy into Eq.~(\ref{eq:52}) we get an analytic expression for the function $\tilde{t}$
\begin{align}\label{eq:65}
\tilde{t}(p)=\frac{x_p^{\sigma}}{1-x_p},
\end{align}
where the pressure $p$ is absorbed in the quantity
\begin{align*}
x_p=e^{-\beta p}.
\end{align*}
The Gibbs free energy per particle is
\begin{align}\label{eq:65a}
\beta G(p,T)= - \ln\!\left[\frac{x_p^{\sigma}}{1-x_p}\right].
\end{align}
where we used the same notation $G$ as the total Gibbs free energy to make the notation easier. As expected, Eq.~(\ref{eq:65a}) shows that a system of pure hard rods is athermal (where all thermodynamic effects are of entropic). In fact, $\beta G(p,T)=f(\beta pa,\sigma/a)$ where $a$ is the lattice parameter (taken as unit in all expressions). The equation of state inferred from the free energy reads
\begin{equation}\label{eq:67}
\frac{1}{\rho}=\sigma+\frac{x_p}{1-x_p},
\end{equation}
from which we get an explicit expression for the mass density $\tilde{\rho} = \sigma\rho$ as a function of the pressure
\begin{equation}
\tilde{\rho}= \frac{\sigma(1-x_p)}{\sigma(1-x_p) + x_p}.
\end{equation}
Other thermodynamics functions follow straightforwardly using standard thermodynamic relations. Entropy, heat capacity, isothermal compressibility, and coefficient of thermal expansion are given respectively by
\begin{equation}
S/k_B =\frac{x_p\ln x_p}{x_p-1} - \ln(1-x_p),
\end{equation}
\begin{equation}
C_p = \left(\frac{x_p\ln x_p}{x_p-1}\right)^{2},
\end{equation}
\begin{equation}
\chi_T = \frac{\beta x_p}{1-x_p}\frac{1}{x_p+\sigma(1-x_p)},
\end{equation}
and
\begin{equation}
\alpha_V/k_B =\frac{1}{1-x_p}\frac{\beta x_p\ln x_p}{x_p+\sigma(1-x_p)},
\end{equation}
where all the quantities are scaled by particle number $N$. The previous thermodynamic functions can be re-expressed explicitly in terms of the mass density $\tilde{\rho}$ and they are given by
\begin{align}
&\beta G(p,T)\!=\! -\!\ln\!\!\left[1+\frac{\sigma(\tilde{\rho}-1)}{\tilde{\rho}}\!\right]\!+\!\sigma\ln\!\!\left[1+\frac{\tilde{\rho}}{\sigma(1-\tilde{\rho})}\right],\\
&\beta p = \ln\!\!\left[1+\frac{\tilde{\rho}}{\sigma(1-\tilde{\rho})}\right],\\
&\frac{S}{k_B} \!\!=\!\!\ln\!\!\left[\!1\!+\!\frac{\sigma(\tilde{\rho}-1)}{\tilde{\rho}}\!\right]\!\!-\!\!\left[\!1\!+\!\frac{\sigma(\tilde{\rho}-1)}{\tilde{\rho}}\!\right]\!\!\ln\!\!\left[\!1\!+\!\frac{\tilde{\rho}}{\sigma(1-\tilde{\rho})}\!\right],\\
&C_p=\left(\frac{\sigma\left[1-\tilde{\rho}\right]}{\tilde{\rho}}\ln\!\!\left[1+\frac{\tilde{\rho}}{\sigma(1-\tilde{\rho})}\right]\right)^{2},\\
&\chi_T = \frac{\beta\sigma}{\tilde{\rho}}\left[1-\tilde{\rho}\right]^2\left[1+\frac{\tilde{\rho}}{\sigma(1-\tilde{\rho})}\right],\\
&\frac{\alpha_V}{k_B} \!=\! \frac{\beta\sigma}{\tilde{\rho}}\!\!\left[1-\tilde{\rho}\right]^2\!\!\left[1+\frac{\tilde{\rho}}{\sigma(1-\tilde{\rho})}\right]\!\!\ln\!\!\left[1+\frac{\tilde{\rho}}{\sigma(1-\tilde{\rho})}\right].
\end{align}
Variations of the different thermodynamic functions as a function of mass density are depicted in Fig.~(\ref{fig:fig1}) for different values of the rod length.
The free energy $G$ increases monotonically with increasing the mass density reflecting the absence of phase transition, which is a well-known result for $1D$ systems with short-range interactions. At low density, increasing the length of the rods at a given volume reduces the free volume accessible for the rods, and hence the free energy decreases. At high density, when increasing the length of the rods, the correlation between rods increases and this rises the free energy. At low density, the pressure rises linearly with temperature as it does in the
classical ideal gas. At high density, the correlation increases, and the variation of the pressure deviates from the ideal gas pressure profiles.
Increasing the density (or decreasing the length of the rods) enhances the ordering and accordingly the entropy $S$ decreases with both increasing the mass density or decreasing the rod length. This has a direct and similar consequence on the heat capacity $C_p$. Both quantities reflect the athermal nature of the hard-core interactions. Isothermal compressibility $\chi_T$ and thermal expansion coefficient $\alpha_V$ express respectively the relative change in volume due to change in pressure (at fixed temperature) or change in temperature (at fixed pressure). Panels for compressibility and coefficient of thermal expansion show the lattice fluid of hard rods is very compressible at low density but at a high density where the system is filled, the volume changes slightly with the applied pressure.
\begin{figure}[htb]
  \begin{center}
 \includegraphics[scale=0.5]{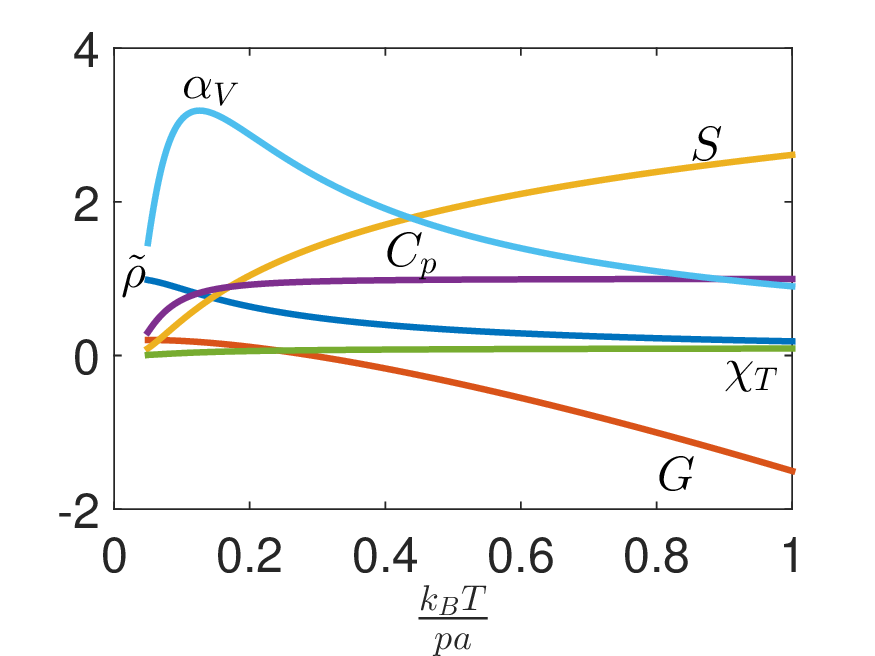}
\end{center}
\caption{Variation of the different thermodynamic functions versus $k_BT/pa$ for rods with pure hard-core interaction.}
  \label{fig:fig2}
\end{figure}
Because a system of pure hard rods is athermal, thermodynamic functions depend only on $\bar{T}=k_BT/pa$. Their variations are shown in Fig.~(\ref{fig:fig2}). Free energy decreases with increasing $\bar{T}$ as the latter disperse the rods and reduce correlations. Rising the quantity $\bar{T}$ also reduces order and hence entropy increases as can be seen in Fig.~(\ref{fig:fig2}).
\subsection{Attractive gravitational interaction}
System of hard rods in continuum with gravitational attraction has been investigated in the different ensemble in the context of studying gravitational collapse \cite{Kumar/Miller:2017,Champion:2015,Youngkins/Miller:2000}. Analytical expressions of the partition function and some other thermodynamic quantities have been worked out in
\cite{Champion:2015,Jang/etal:2017}. Here, we consider the lattice version of this system where the rods are distributed on a lattice of L sites. The rods interact through an attractive two-body gravitational potential of the form
\begin{align}
H = \sum\limits_{i<j}\phi_{i,j}
\end{align}
with
\begin{align}\label{eq:lenard_energy}
\phi_{i,j} = \left\lbrace\begin{array}{ll}
\infty& \mbox{if $|z_i - z_j|<\sigma$ }\\
gm^2|z_i - z_j| & \mbox{if $|z_i - z_j|\geq \sigma$ }
\end{array}\right.
\end{align}
where $m$ is the mass of the rod, $g$ is the strength of gravitation and $z_i$ is the position of rod $i$. The total potential energy can be rewritten as
\begin{align}\label{eq:transf}
\phi(z_1,z_2,\ldots,z_N)&=\sum_{1\leq i<j\leq N}\phi_{i,j}\nonumber\\
&=\sum_{i=1}^{N-1}a_i(z_{i+1}-z_i)
\end{align}
with
\begin{align}
a_i=gm^2i(N-i)
\end{align}
To calculate the partition function, we make a variable change in Eq.~(\ref{eq:transf}) where we switch from using the position of the rods $z_i$ as coordinates to the distance between the rods and the center of gravity as a new coordinates. Inserting the total interaction energy into Eq.~(\ref{eq:52}) we get an exact expression for  the isobaric partition function
\begin{align}
\mathcal{Z}=\prod_{i=0}^N\tilde{t}_i,
\end{align}
where 
\begin{align}
\tilde{t}_i=\frac{\omega_i^{\sigma}}{1-\omega_i}, \hspace{10mm}\omega_i=e^{-\beta(p+a_i)}.
\end{align}
The partition function can be simplified into
\begin{align}\label{eq:pf_gravity_lat}
\mathcal{Z}=e^{-\beta p_0\sigma}\prod_{i=1}^{N}\frac{1}{1-\omega_i}
\end{align}
with $p_0=[p(N+1)+gm^2N(N^2-1)/6]$. Thermodynamics functions inferred from Eq.~(\ref{eq:pf_gravity_lat}) are given by
\begin{figure}[t]
  \begin{center}
 \includegraphics[width=42mm]{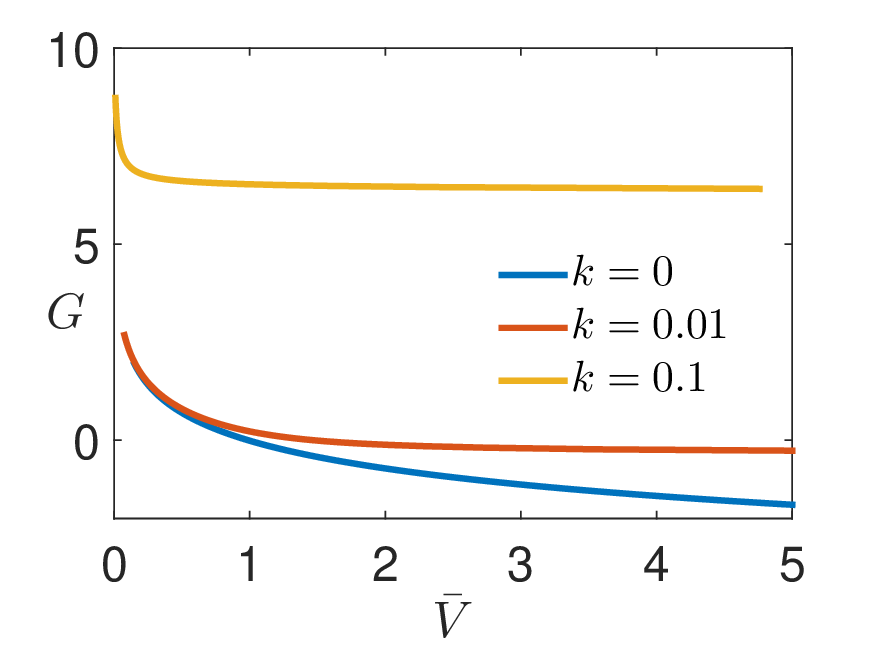}
 \includegraphics[width=42mm]{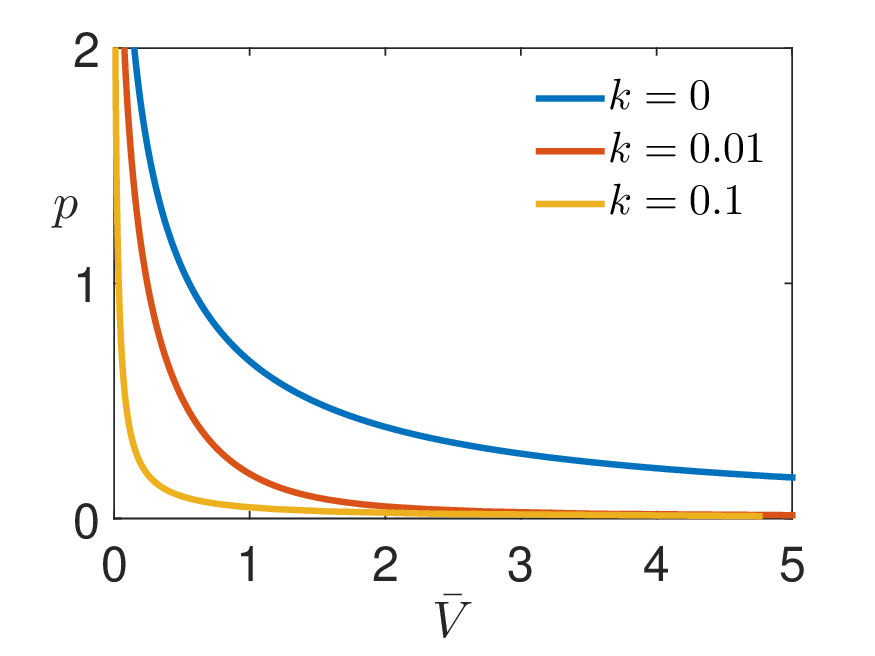}\\
 \includegraphics[width=42mm]{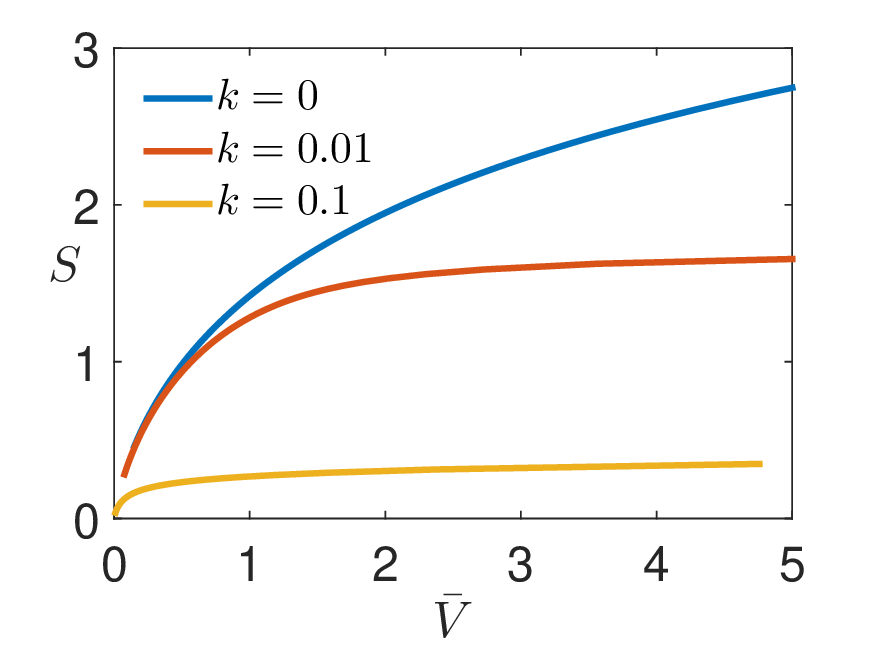}
 \includegraphics[width=42mm]{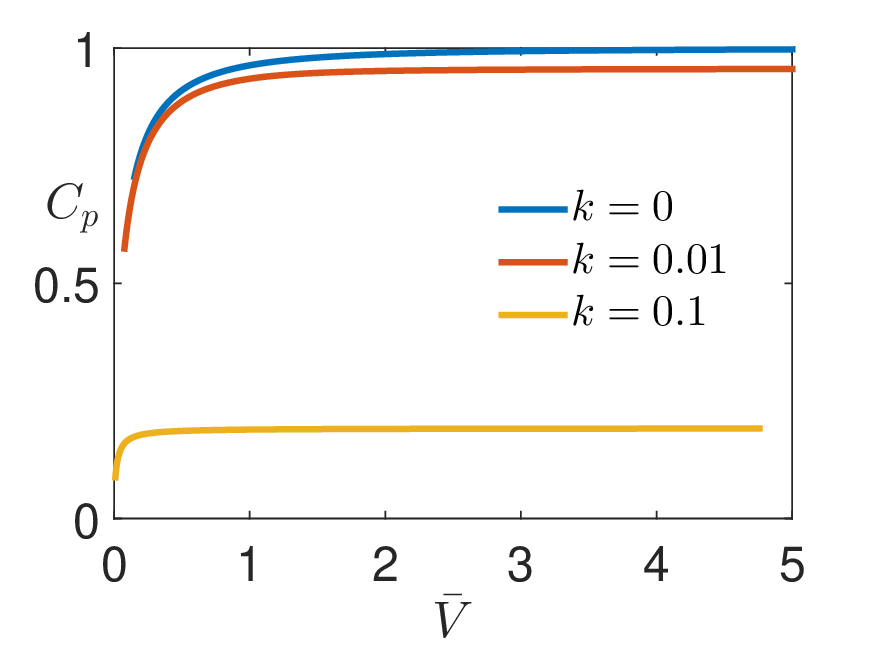}\\
 \includegraphics[width=42mm]{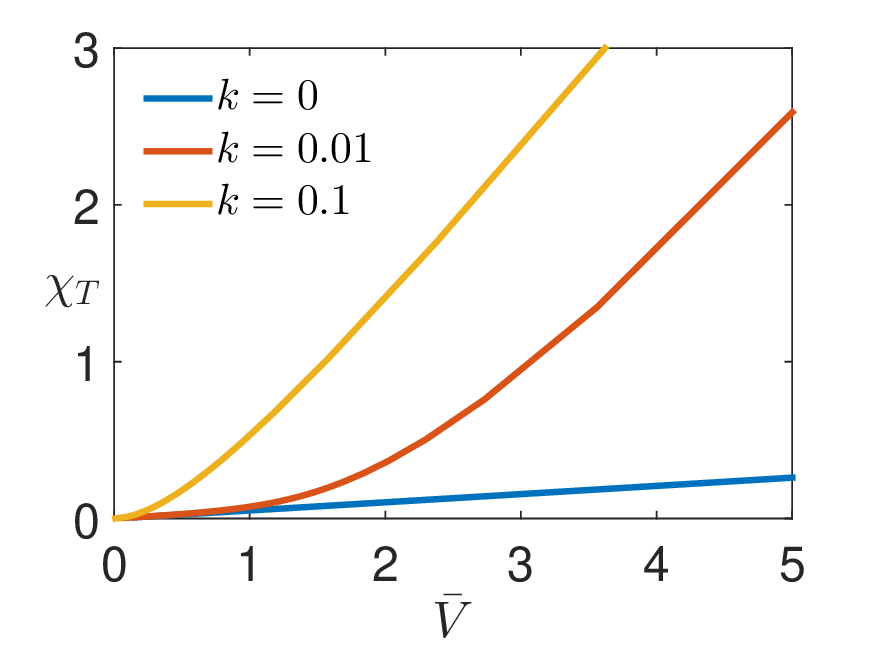}
 \includegraphics[width=42mm]{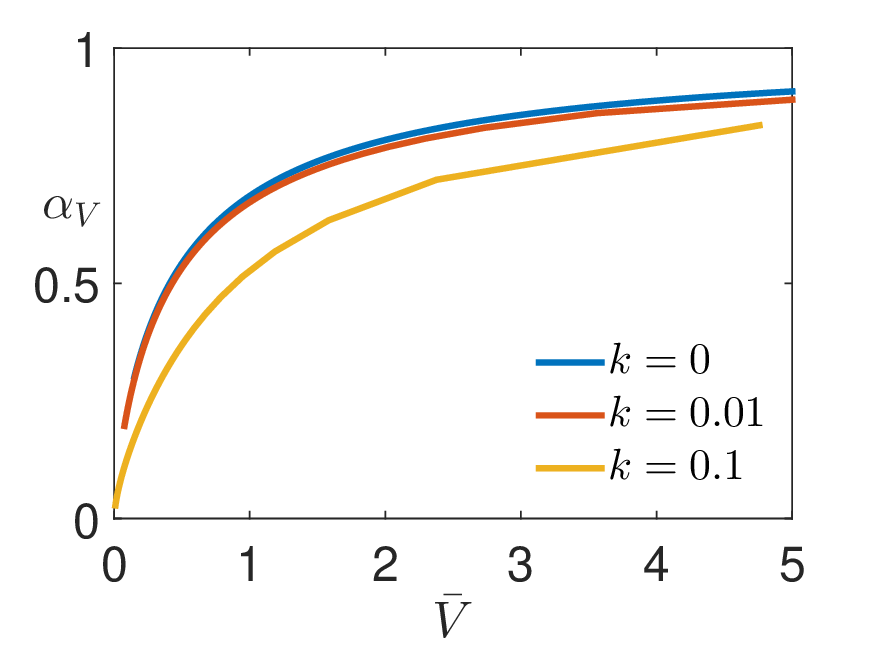}
\end{center}
\caption{Free energy, pressure, entropy, heat capacity, compressibility, and coefficient of thermal expansion versus excess volume $\bar{V}$ of a gas of hard rods on a lattice with attractive gravitational interactions. The parameter $k$ is $k=\beta m^2g$. The rod length is fixed at $\sigma=5$.}\label{fig:fig3}
\end{figure}
\begin{align}
&\beta G=\beta p_0\sigma + \sum_{i=1}^{N}\ln(1-\omega_i)\label{eq:gr_lt1},\\
&\frac{1}{\rho}=\sigma +\frac{1}{N+1}\sum_{i=1}^{N}\tilde{\omega}_i\label{eq:gr_lt2},\\
&\mathcal{S}=-\sum_{i=1}^{N}\ln(1-\omega_i) + \beta\sum_{i=1}^{N}\tilde{\omega}_i(p+a_i)\label{eq:gr_lt3},\\
&C_p=\beta^2\sum_{i=1}^{N}(p+a_i)^{2}\frac{\tilde{\omega}_i^2}{\omega_i}\label{eq:gr_lt4},\\
&\chi_T=\frac{\beta\rho}{N+1}\sum_{i=1}^{N}\frac{\tilde{\omega}_i^2}{\omega_i}\label{eq:gr_lt5},\\
&\frac{\alpha_V}{k_B}=\frac{\beta^2\rho}{N+1}\sum_{i=1}^{N}(p+a_i)\frac{\tilde{\omega}_i^2}{\omega_i}\label{eq:gr_lt6},
\end{align}
where
\begin{align}
\tilde{\omega}_i=\frac{1}{\omega_i^{-1}-1}.
\end{align}
We have evaluated Eq.~(\ref{eq:gr_lt1})-(\ref{eq:gr_lt6})  numerically using the Broyden algorithm and they are presented in Fig.~(\ref{fig:fig3}) for different values of the parameter $k=\beta gm^2$, that expresses the ratio of the gravitational energy to thermal energy. For convenience, we have shown variations of all thermodynamic quantities as functions of the excess volume
\begin{align}
\bar{V}=\frac{1}{\rho}-1,
\end{align}
\begin{figure}[htb]
  \begin{center}
 \includegraphics[scale=0.5]{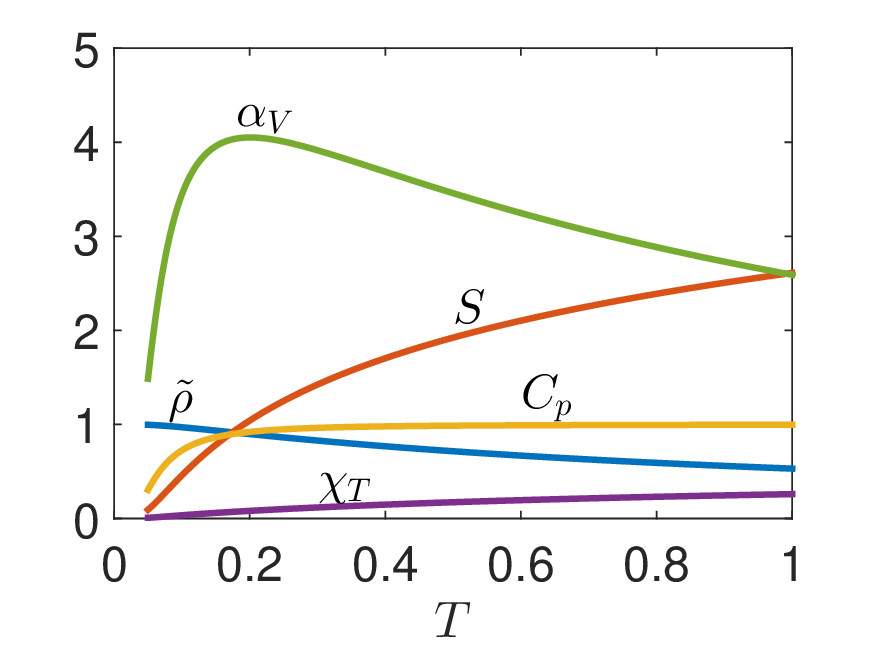}
\end{center}
\caption{Variation of the different thermodynamic functions versus temperature for rods with gravitational interaction. The number of particles, pressure and  the strength of the gravitational are set respectively to $N=50$, $p=0.2$ and $gm^2=1.0$. The rod length is fixed at $\sigma=5$.}
  \label{fig:fig4}
\end{figure}
in which the lattice parameter is set to one ($a=1$). Because of the finite size of the system, no phase transition happens as shown in Fig.~(\ref{fig:fig3}).
As expected, the free energy $G$ increases with increasing the strength of the gravitational attraction, and this change is more pronounced at low values of the excess volume where the rods are close to each other. This has its effect on the pressure $p$ in which the pressure increases (decreases) when decreasing (increasing) the excess volume. At low excess volume and due to the long-range attraction, any small change in the excess volume strongly affects the pressure and vice versa. In kinetic theory, the distinction is made between kinetic pressure and interaction pressure. Kinetic pressure is attributed to elastic collisions between rods and container walls. Attractive interatomic couplings tend to reduce the total pressure. The entropy $S$ decreases with increasing the attractive forces due to the fact the latter produces more order in the system. At very strong attraction, all the rods stick together to form a larger cluster and the entropy is almost zero. Contrary to the athermal hard-core interactions, in the case of attractive interactions, the temperature is involved explicitly in the expression of entropy and heat capacity $C_p$.
The effect of correlation induced by the presence of attraction reduces drastically the heat capacity and also makes the lattice fluids more sensitive to the effect of reservoir pressure as shown for the curves of compressibility and coefficient of thermal expansion.
Variations of all thermodynamics functions versus temperature are shown in Fig.~(\ref{fig:fig4}), where the monotonic variations confirm the absence of phase transitions. The energy transfer during isobaric expansion is naturally
expressed as a change in enthalpy, $E=U+pV$ (with $U$ being the internal energy). The system performs work on the environment ($\Delta W <0$) if heat is added ($\Delta Q>0$). Energy conservation with $U=0$ implies that $\Delta W=-\Delta Q$ and we have
\begin{align}
\Delta W = -p\Delta V=-\Delta E,\nonumber\\
\Delta Q = \int_{T_{ini}}^{T_{fin}}C_pdT = +\Delta E\nonumber.
\end{align}
\subsection{Charged hard rods with hard wall}
System of charged particles with charged boundaries and confined into a one-dimensional channel has been introduced a long time ago as a model for one-component plasma but the model still attracts much interest \cite{Dean/etal:2014,Dubin:1997,King/etal:2002,Heo/etal:2019}. The model has been shown to be very useful in studying the mechanism of charge regulation \cite{Frydel:2019}, ionic liquids near charged interfaces 
\cite{Hayes/etal:2015,Ivanistse/Fedorov:2014,Fedorov/Kornyshev:2014}, and array of charged smectic layers or lipid multilayers \cite{Dean/etal:2014}. 
We apply the general formalism of Sec.~(\ref{sec:sec_3}) to the case of $N$ charged rods in the presence of charged hard wall (considered as a neutralizing background). The Hamiltonian is given by
\begin{figure}[htb]
  \begin{center}
 \includegraphics[width=42mm]{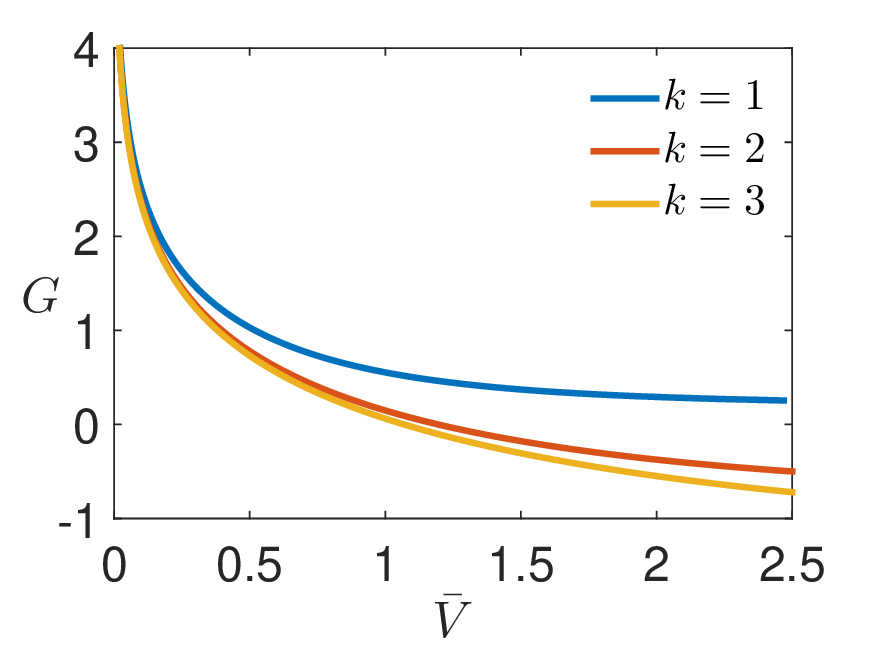}
 \includegraphics[width=42mm]{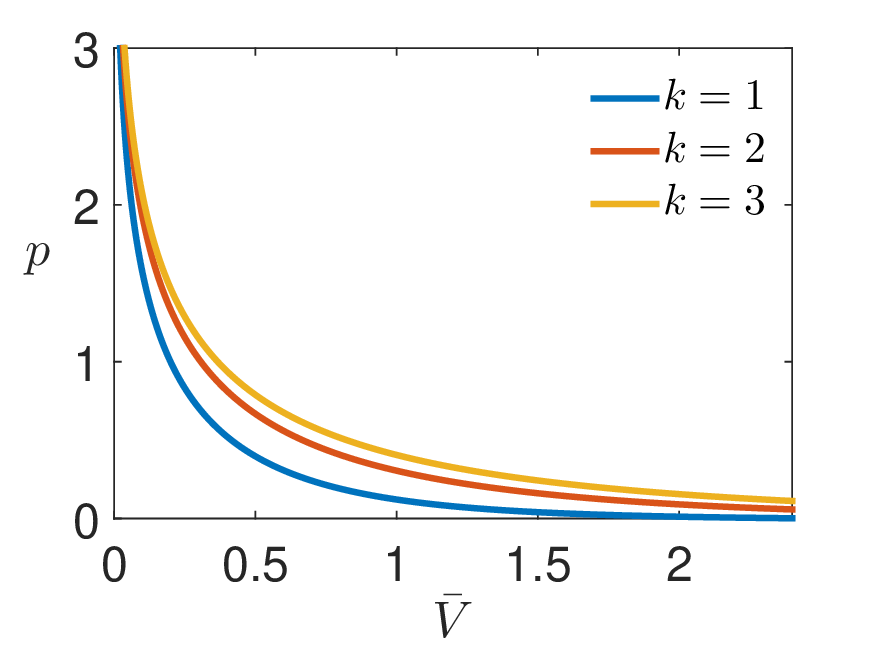}\\
 \includegraphics[width=42mm]{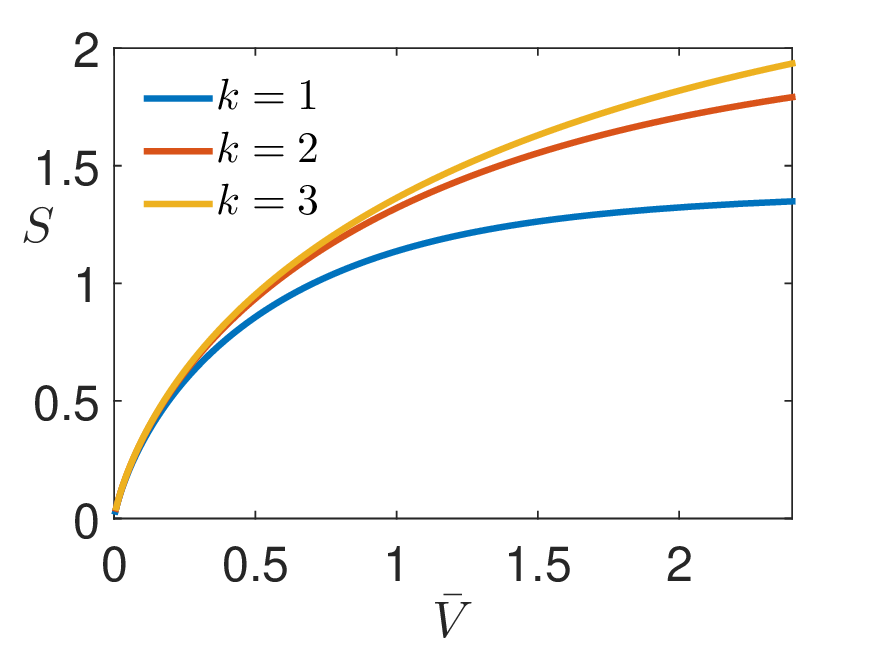}
 \includegraphics[width=42mm]{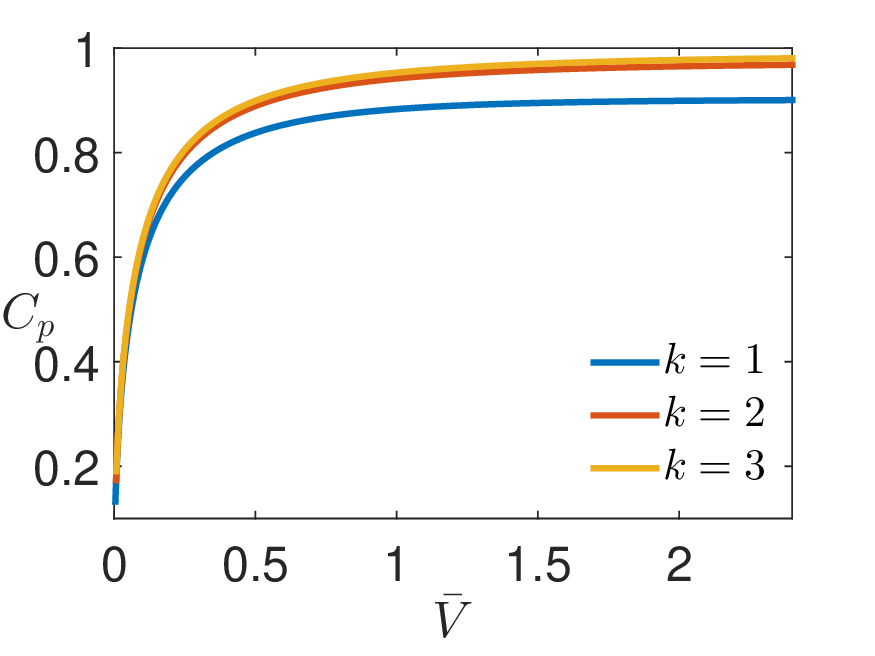}\\
 \includegraphics[width=42mm]{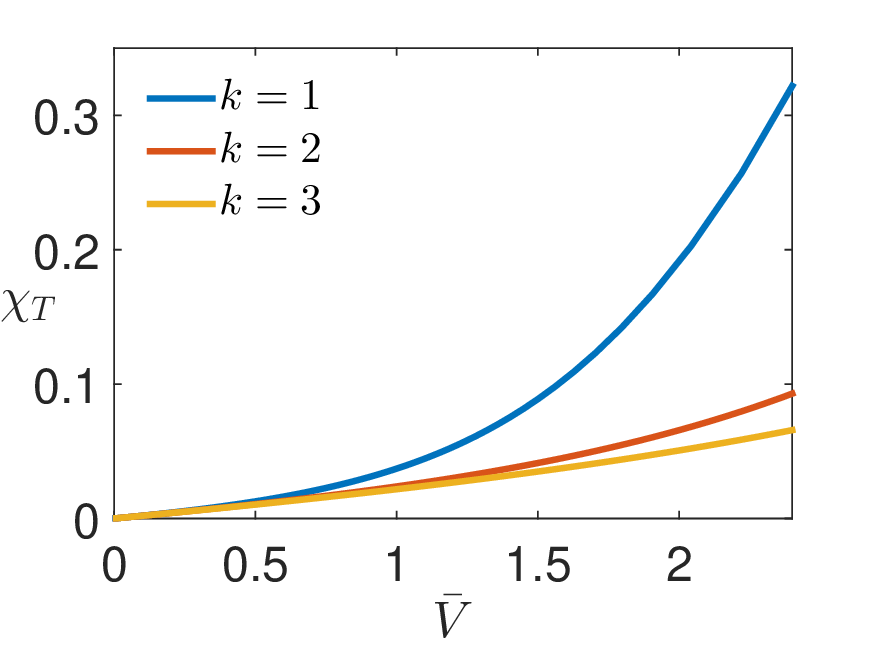}
 \includegraphics[width=42mm]{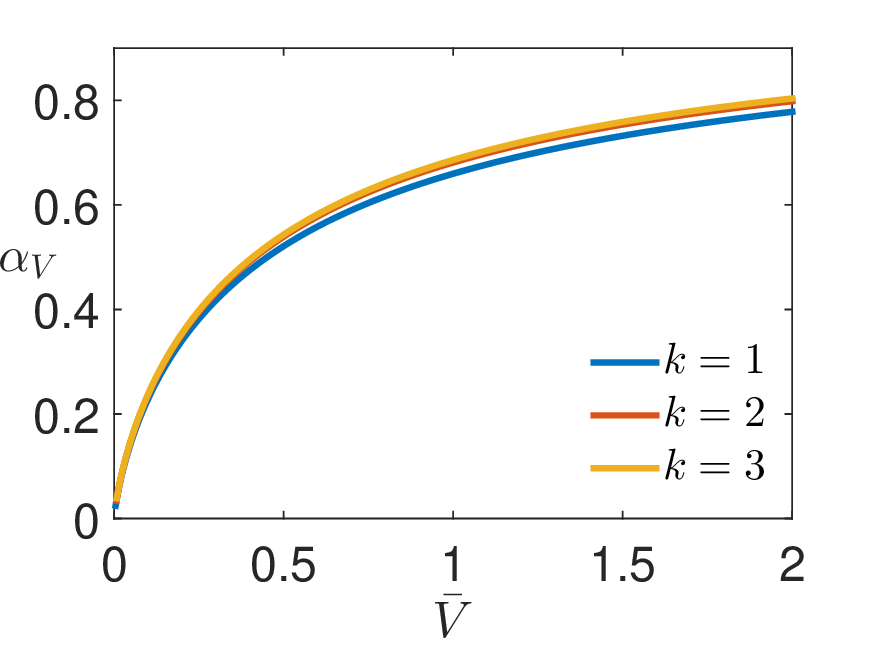}
\end{center}
\caption{Free energy, pressure, entropy, heat capacity, compressibility, and coefficient of thermal expansion versus excess volume $\bar{V}$ of a gas of hard rods with $k=\beta/N\lambda$. The rod length is set to $\sigma=5$.}
  \label{fig:fig5}
\end{figure}
\begin{equation}
H = \sum\limits_{i<j}\phi_{i,j}
\end{equation}
where the one dimensional coulomb interactions between rods have the form
\begin{equation}
\phi_{i,j}=\left\lbrace \begin{array}{ll}
\infty   &   \mbox{for $    |z_i-z_j| < \sigma $} \\
-\frac{|z_i-z_j|}{N\lambda} &   \mbox{for $    |z_i-z_j| \geq \sigma $}
\end{array}\right. 
\end{equation}
with $\lambda$ is the inverse of the interaction strength and the factor $N$ in the denominator has been introduced to neutralize the system \cite{Frydel:2019}. The wall confinement is represented by the potential energy 
\begin{align}
\beta V_{w}(z_i)=\frac{z_i}{\lambda}
\end{align}
applied on particle $i$. The combined effects of Coulomb interparticle interaction and wall potential can be rearranged into \cite{Frydel:2019}
\begin{align}\label{eq:a_charged_rod}
\phi(z)=a_iz; \hspace{10mm}a_i=\frac{2N-2i+1}{N\lambda}
\end{align}
The function $\tilde{t}_i$ is given by
\begin{align}
\tilde{t}_i(p,T)=\frac{e^{-\beta(p+a_i)´\sigma}}{1-e^{-\beta(p+a_i)}}
\end{align} 
and the free energy is given by
\begin{equation}
\beta G(p,T)= \beta \!\left(p(N+1)+\frac{N}{\lambda}\right)\!\sigma+\sum_{i=1}^{N}\ln\!\left(1-\omega_i\right)
\end{equation}
with $\omega_i=e^{-\beta(p+a_i)}$. 
Thermodynamic functions have the same analytical forms as for the gravitational interactions but with a different $\tilde{\omega}$,
\begin{align}
\tilde{\omega}_i=\frac{1}{e^{\beta(p+a_i)}-1}
\end{align} 
where $a_i$ is given in Eq.~(\ref{eq:a_charged_rod}).
 Numerical results for the thermodynamic functions are shown in Fig.~(\ref{fig:fig5}) for different values of the strength of the electrostatic interaction $\lambda$ encoded in the parameter $k=\beta/N\lambda$. The pressure results from the ideal gas contribution and from the interaction. Contrary to the attractive gravitational interactions, the repulsive electrostatic interaction augments the pressure as shown in Fig.~(\ref{fig:fig5}). Compressing the gas of lattice hard rods means removing vacant cells. Removing vacant cells lowers the entropy. Lowering the entropy at $T > 0$ (Fig.~(\ref{fig:fig6})) implies that heat is expelled. The heat capacity $C_p$ rises exponentially from zero  both at low $\bar{V}$ (Fig.~(\ref{fig:fig5})) or $T$ (Fig.~(\ref{fig:fig6})) and approaches unity asymptotically as $\bar{V}$ or $T$ becomes large. The combined effect of electrostatic interaction and charged hard wall makes the lattice hard rods more sensitive to external pressure and less sensitive to the effect of temperature as shown for the panels of compressibility and coefficient of thermal expansion in Fig.~(\ref{fig:fig5}).
\begin{figure}[t]
  \begin{center}
 \includegraphics[scale=0.5]{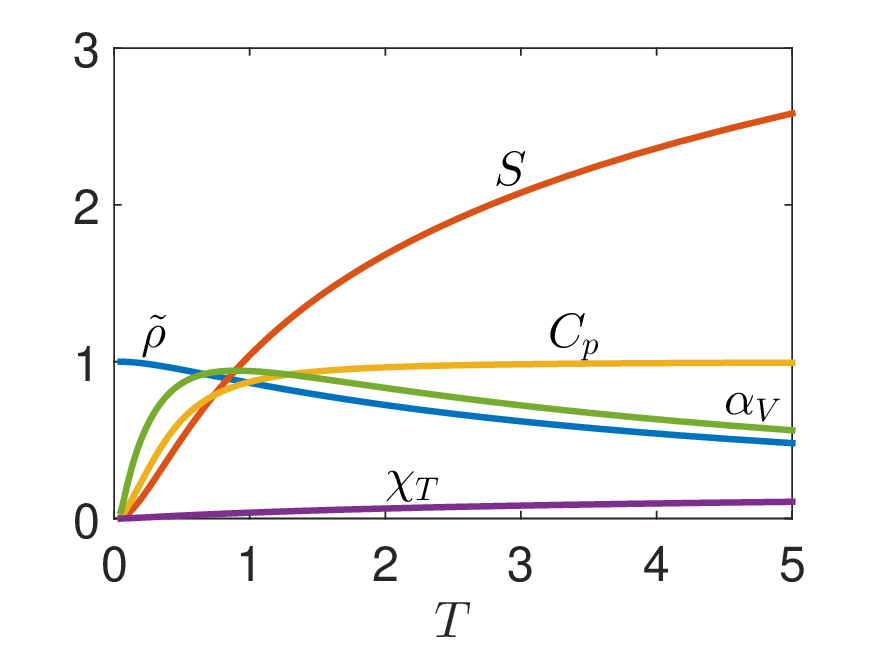}
\end{center}
\caption{Variation of the different thermodynamic functions versus temperature for charger hard rods in neutralizing background and under sem-confinement. The number of particles, pressure and  the strength of the gravitational are set respectively to $N=50$, $p=0.2$, $\sigma=5$ and $\lambda=1.0$.}
  \label{fig:fig6}
\end{figure}
\section{Conclusion}
In this paper, we have carried out exact analytic and numerical studies of some statistical lattice models in the isobaric ensemble. Despite that the quantum description is more natural at the micro and nanoscale, it has been proved that for the system at high temperature where thermal fluctuation is compared or dominated over the quantum fluctuation, the classical description is appropriate and highly justified. First, we have established a general formalism to calculate the exact partition function and the corresponding thermodynamics quantities for arbitrary nearest-neighbor interactions. The calculation is made for the Isobaric ensemble, but an inverse Laplace transform can be used to get the canonical partition function. However, in the thermodynamic limit, the thermodynamic average in the isobaric partition function agrees to $1/N$ with those of the canonical partition function \cite{Lebowitz/Percus:1961}.  Applications are made to systems with a hard-core exclusion, gravitational attractive interactions, and charged hard rods in semi-confinement with one hard wall. Our study is motivated by the recent extensive experimental study of the one-dimensional atomic chains and their formation on the surface. Extension of this to the case of an inhomogeneous system with an arbitrary external field currently works under progress. Another possible extension of this work is to study phase transitions in systems with long-range interactions, such as gravitational collapse \cite{Bakhti/etal:2018,Bakhti:2021}. These phase transitions have been shown to have an experimental evidence even for one dimensional systems \cite{Chalony/etal:2013}.
For these systems, one needs to check the stabilities (existence of equilibrium states) in the thermodynamic limit.
To the best of our knowledge, stabilities of such systems have been well studied
in the canonical and microcanonical ensembles but not in the isobaric ensemble. This due to the fact that the canonical and microcanonical ensemble are the more relevant and natural ensembles to study phase transition
in the presence of long-range gravitational force (in which the theory
fit very well with observations). In addition, due to the non-extensive nature of the long-range gravitation force, the phase transition can happen for the same system in an ensemble but not in another. So, to use the present results, one needs first to perform a Laplace transform of the isobaric partition function to get the canonical
partition function, then use standard thermodynamic analysis to determine the phase diagram.

\begin{acknowledgements}
B.~B. would like to thank Prof. Gerhard M\"uller for very valuable discussion. The authors gratefully thank Prof. Siegfried Dietrich for insightful suggestion of some useful references and Anke Geigle for sending the papers.
\end{acknowledgements}

\section*{Compliance with Ethical Standards}
The authors declare the following competing interests:\\
\textbf{Funding}: R.~C. acknowledges the financial support received from Karlsruhe Institute of Technology (KIT).\\
\textbf{Conflict of Interest}: The authors declare no competing interests.\\
\textbf{Ethical Conduct}: This material is the authors' own original work, which has not been previously published elsewhere.
\bibliographystyle{apsrev4-2} 
\bibliography{dft}

\begin{thebibliography}{58}
\bibitem [{com()}]{comm:kinetic_part_function}%
  \BibitemOpen
  \href@noop {} {}\bibinfo {note} {For system with N particles, the total
  partition function can be factorized as $Z=Z_{kin}Z_{pot}$, where
  $Z_{kin}=\int d^{3}p e^{-\beta p^{2}/2m}$ ($d^{3}p=d^{3}p_1\ldots d^{3}p_N$).
  The kinetic part of the partition function $Z_{kin}$ can be integrated (or
  summed in the case of lattice system) exactly and it yields a constant term
  proportional to the de Broglie thermal wave length. As a result, an extra
  additive term appears in the free energy, which does not affect the
  thermodynamics.}\BibitemShut {Stop}%
\end{thebibliography}

\end{document}